\documentclass[10pt,twoside,twocolumn]{article}
\usepackage{amsmath}
\usepackage{amsfonts}
\usepackage{amssymb}
\usepackage{graphicx}
\title{\textbf{Molecular Dynamics in Dissociative Electron Attachment to CO probed by Velocity Slice Imaging}}
\author{\textbf{Pamir Nag$^1$} and \textbf{Dhananjay Nandi$^2$}\\ Indian Institute of Science Education and Research Kolkata, Mohanpur 741246, India\\ \small{email: $^1$pamir1118@iiserkol.ac.in, $^2$dhananjay@iiserkol.ac.in}}

\date{}
\begin{document}
\twocolumn[
  \begin{@twocolumnfalse}
    \maketitle
    \begin{abstract}
      Kinetic energy and angular distributions of O$^-$ ions formed by dissociative electron attachment to CO molecule have been studied for 9, 9.5, 10, 10.5, 11, 11.5 eV incident electron energies around the resonance using time sliced velocity map imaging spectrometer. Detailed observations clearly show two separate DEA reactions lead to the formation of O$^{-}$ ions in the ground $^{2}P$ state along with the neutral C atoms in ground $^{3}P$ state and first excited $^{1}D$ state, respectively. Within the axial recoil approximation and involving four partial waves, our angular distribution results clearly indicate that the two reactions leading to O$^{-}$ formation proceed through the distinct resonant state(s). For the first process, more than one intermediate states are involved. Whereas, for the second process, only one state is involved. The observed forward-backward asymmetry is explained due to the interference between the different partial waves that are involved in the processes. 
    \end{abstract}
  \end{@twocolumnfalse}
]

\section{Introduction}
Low energy electron-molecule collision leading to dissociative electron attachment (DEA) is an important process from the fundamental as well as the application point of view. DEA study of molecules are very important starting from astrophysics to biology. The resonance formation can be used for a single- or double-strand break in DNA \cite{DNA:science}. Site specific fragmentation \cite{prl:site_DN} can also lead to selective bond cleavage in DNA \cite{PhysRevLett.110.023201}. Chandler and Houston \cite{ar:chandler} first used the imaging technique to study the molecular dynamics. Later, velocity map imaging (VMI) technique \cite{ar:parker} and slice imaging \cite{ar:slice_ima,ar:dc_slice,ar:ashfold} helped to study the angular distribution and kinetic energy distribution simultaneously and very accurately in the photo-dissociation dynamics. Recently, this method has been modified and implemented in the low energy electron molecule collision studies \cite{inst:DN} for the first time. Since then, the velocity slice imaging (VSI) technique in its various forms have been successfully employed to study the low energy electron-molecule collisions by different groups \cite{inst:adaniya,inst:moradmand,inst:china1,inst:china2} in the recent times. Very recently, we developed a modified velocity slice imaging spectrometer to study the low and intermediate energy electron-molecule collision experiments. In this study, the spectrometer has been probed to measure the kinetic energy and angular distribution of O$^-$ ions produced from CO by DEA process.
\begin{figure}[h]
\centering
\includegraphics[scale=.45]{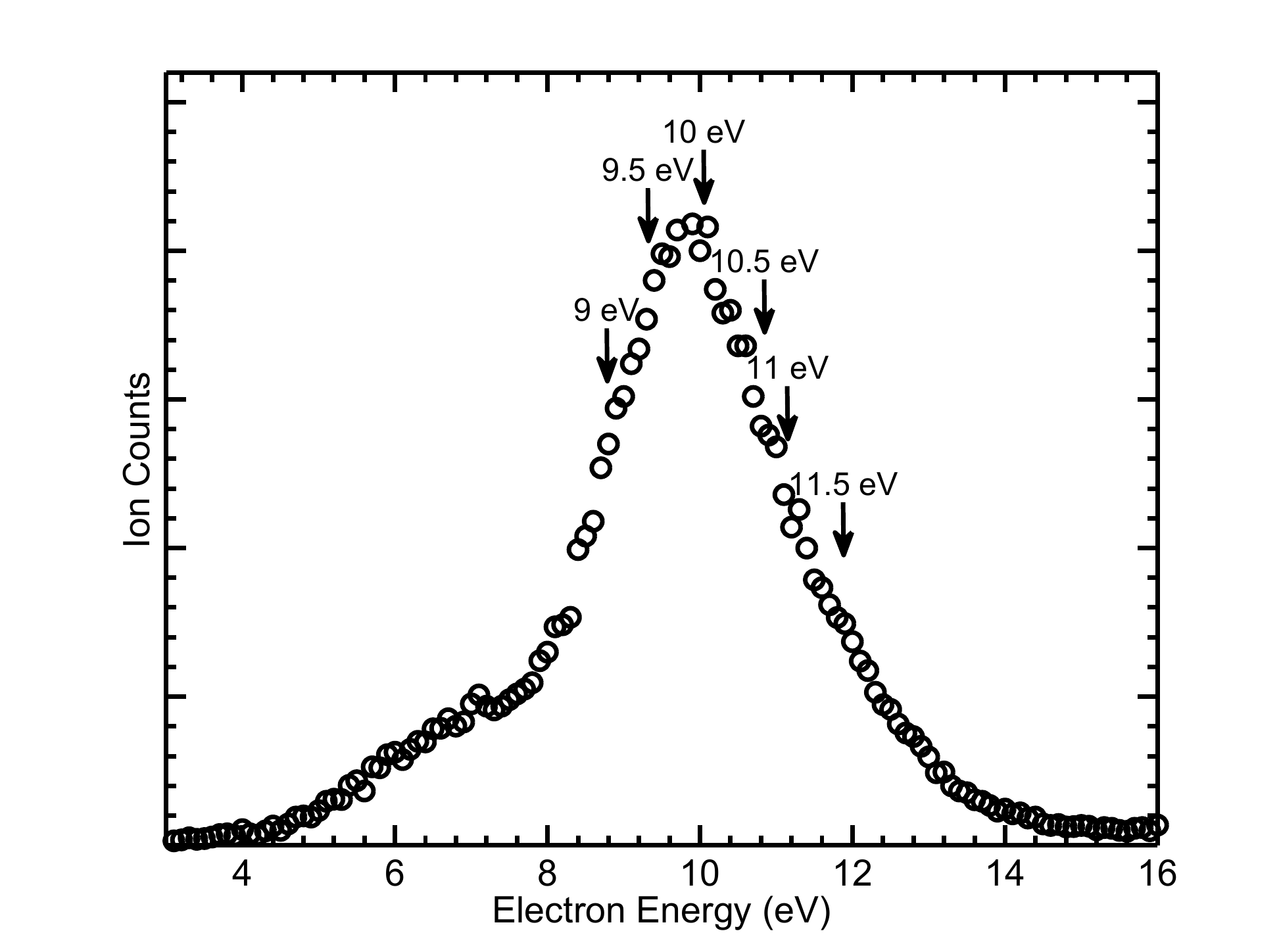} \caption{\small{Ion yield curve of O$^-$ produced from DEA to CO. The arrows indicate the energies at which the images are taken.}} \label{fig:ion_yield}
\end{figure}

The O$^-$ ion formation from CO due to electron impact was first observed by Vaughan \cite{co:vaughan} back in 1931. Rapp and Briglia \cite{ref:rapp} measured the absolute cross section and reported to observe the dissociative electron attachment peak near 9.9 eV. The dominant process leading to the O$^-$ formation (Process I) is $$e^- +\text{CO}(^1\Sigma^+)  \rightarrow \text{CO}^{-*} \rightarrow \text{O}^- (^2P) + \text{C} (^3P) .$$ Through energy analysis of the ions by Chantry \cite{co:chantry}, proposed a second process for the O$^-$ formation (Process II) as: $$e^- + \text{CO}(^1\Sigma^+) \rightarrow \text{CO}^{-*} \rightarrow \text{O}^- (^2P) + \text{C}^* (^1D).$$ 
Hall {\it et al.} \cite{co:hall} measured the kinetic energy distribution of the O$^-$ ions at three specific angles, and also the angular distribution of the ions and proposed the intermediate state might be a $\Pi$ state. Morgan {\it et al.} \cite{co:tennyson} recently computed the potential energy curve of the neutral CO molecule and the resonance states using R-matrix formalism. Tian {\it et al.} \cite{co:xi} recently studied the angular distribution of O$^-$ ion from CO due to DEA using velocity slice imaging (VSI) and proposed the presence of coherent interference between the different states that are involved. In this article, we report the kinetic energy distribution of the negative ions over a broad incident electron energy range of 9 eV to 11.5 eV around the resonance and also the angular distribution of the O$^-$ ions depending on their kinetic energy distributions for the above mentioned electron energy range.

\begin{figure}
\centering
\includegraphics[scale=.45]{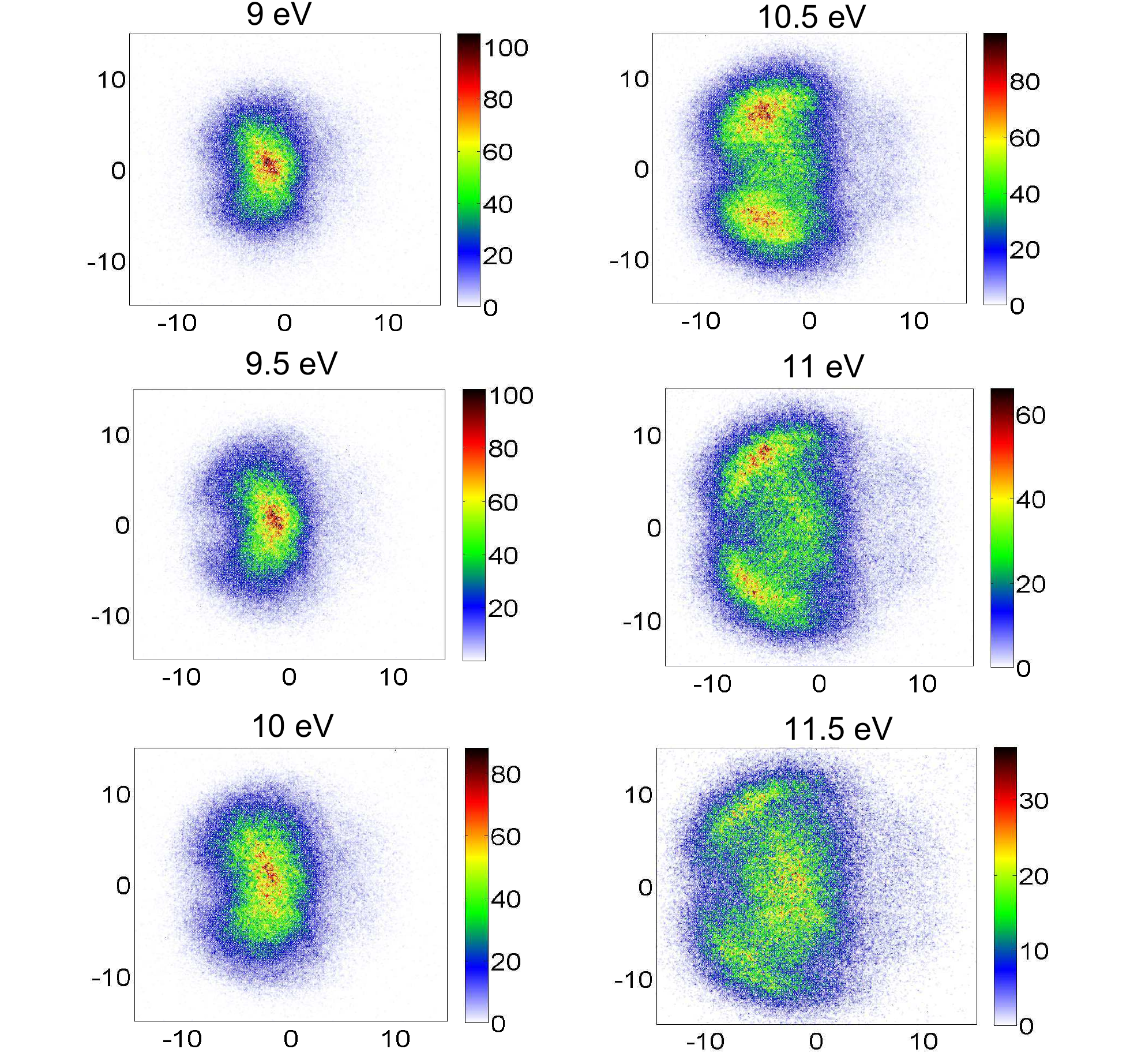} \caption{\small{Time sliced images at different incident electron energies. The incident electron beam direction is along the horizontal axis from left to right through the center of each image.}} \label{fig:3d_pics}
\end{figure}
\begin{figure}
\centering
\includegraphics[scale=.45]{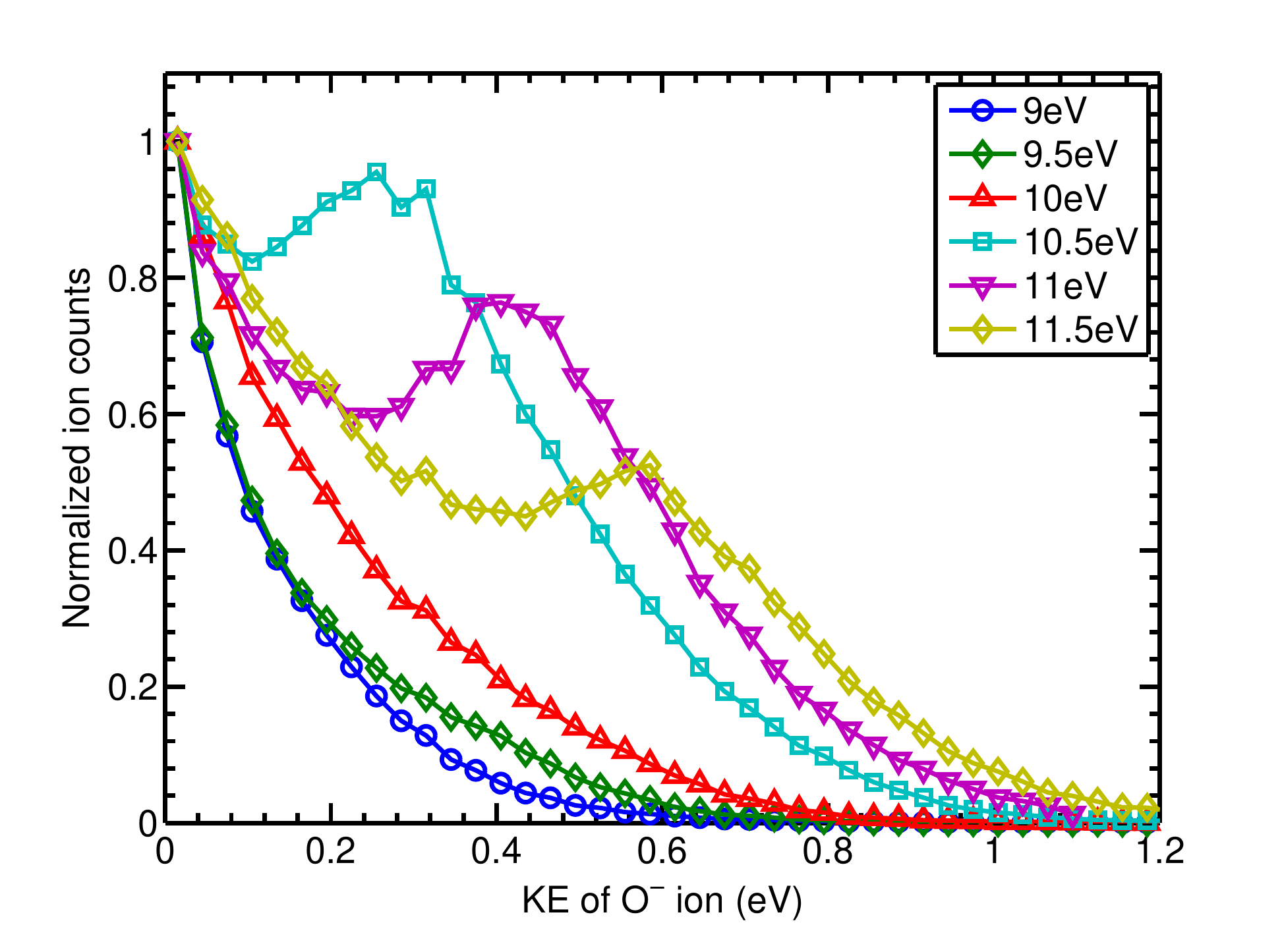} \caption{\small{KER of O$^-$ ion at different incident electron energies.}} \label{fig:ke_dist}
\end{figure}
\section{Instrumentation}
Negative ions are formed due to low energy electron capture and subsequent dissociation. The measurements are performed under high vacuum condition at the base pressure below $\sim$ 10$^{-8}$ mbar. A magnetically well collimated pulsed electron beam of 200 ns duration, 10 kHz repetition rate and with controlled energy is passed through the interaction region where it interacts with an effusive molecular beam produced by a capillary tube. The molecular beam is directed towards the detector and along the axis of the spectrometer. We have used a custom build electron gun consisting of thermally heated filament with typical resolution of 0.5 eV. The magnetic field used to collimate the electron beam is about 40 Gauss. A pair of magnetic coil (Helmholtz type) is mounted outside the vacuum chamber to produce the uniform magnetic field at the interaction region. After it has passed, a negative pulsed extraction field is applied and the negative ions are extracted from the source region into the VMI spectrometer. The extraction pulse duration used in the present experiment is 4 $\mu$s and is applied 100 ns after the electron gun pulse. The delayed extraction provide appropriate time spread for better time sliced image. The VMI spectrometer is like a three field time-of-flight spectrometer \cite{inst:DN} which focuses ions starting from a finite volume onto a two-dimensional position sensitive detector such that ions with a given velocity are mapped to a point on the detector irrespective of their spatial location in the source region. The two-dimensional position sensitive detector consists of three micro channel plates (MCPs) in Z-stack configuration and a three layers delay line hexanode \cite{hex1}. The time-of-flight (ToF) of the detected ions is determined from the back MCP signal whereas the x and y positions of each detected ions are calculated from the three anode layer \cite{hex1} placed behind the MCPs. The x and y position along with ToF of each detected particles are acquired and stored in a list-mode format (LMF) using the CoboldPC software from RoentDek. The central slice through the `Newton Sphere' contains the full angular and translational energy information. The central sliced image is obtained by selecting appropriate time window during the off-line analysis from the stored LMF file using the CoboldPC. Such time sliced image corresponds to the ions ejected in the plane parallel to the detector containing the electron beam axis. 

The typical FWHM of the ToF of the O$^-$ ions produced in this energy range is about 250 ns. We have taken a 50 ns time sliced image from the central part of the entire Newton Sphere. The complete information about the kinetic energy release and angular distribution of the negative ions can be obtained from this central slice. For incident electron beam energy calibration we have considered the O$^-$/CO resonance peak (shown in figure~\ref{fig:ion_yield}) to be at 9.9 eV \cite{ref:rapp}. To measure the kinetic energy release (KER) of the negative ions we have calibrated our system using the energy release of O$^-$/O$_2$ at 6.5 eV \cite{o2:dn_cross}. We also have cross-checked the  kinetic energy calibration by measuring the kinetic energy of O$^-$ produced by electron attachment at 8.2 eV of CO$_2$ \cite{co2:slaughter}.

To get the ion yield curve a different set of data acquisition system has been used. For this purpose the signal from MCP only has been taken. The MCP signal is amplified through a Fast Amp and then fed to a Constant Fraction Discriminator (CFD). The output from CFD is fed to STOP of a Nuclear Instrumentation Module (NIM) standard Time-to-Amplitude Converter (TAC) and START is generated from the master pulse used in the electron gun. The output of the TAC is connected to a Multichannel Analyser (MCA, Ortec model ASPEC-927) and finally communicated with the data acquisition system installed in a dedicated computer via high-speed USB 2.0 (Universal Serial Bus) interface. A home made LabVIEW based data acquisition system has been used to get the ion yield curve. Using this software at first the ToF has been obtain, then by selecting only the channel corresponding to a particular mass the electron energy versus the number of ions produced have been measured.
\begin{figure}
\centering
\includegraphics[scale=.38]{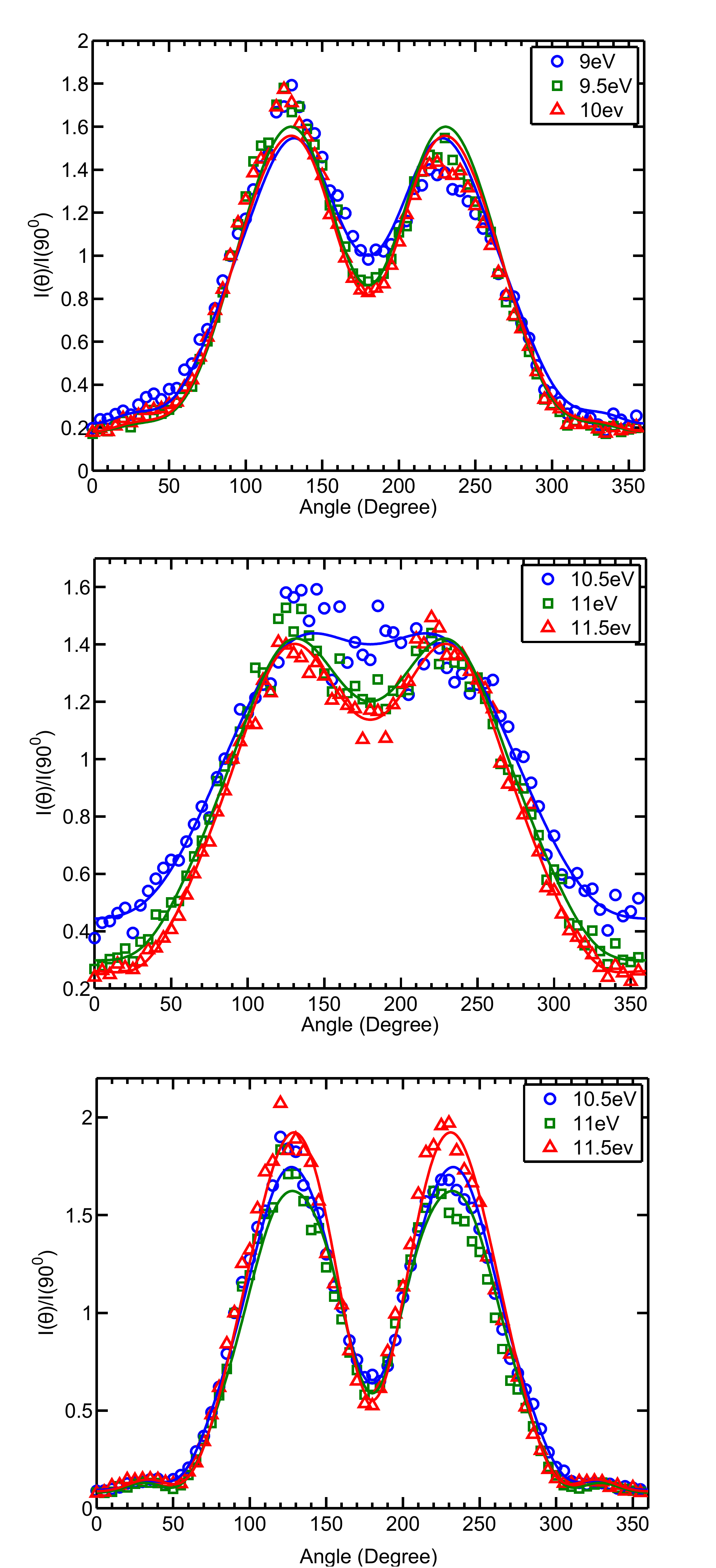}  \caption{\small{The fitted angular distribution for different incident electron energies. Angular distribution for entire KER ranges shown in top. Middle one shows the angular distribution of the ions with KER in between 0 eV to the respective first minima in KER distribution curve(fig~\ref{fig:ke_dist}). Bottom image is the angular distribution for the KER between the respective first minima to the maximum KER.}} \label{fig:ang_dist}
\end{figure}
\begin{table*}
\centering
\caption{Fitting parameters for the angular distribution of the O$^{-}$ ions taken at 9, 9.5 and 10 eV incident electron energies. The angular distributions are fitted with \small{$\Sigma$ to $\Sigma$ and $\Pi$ transition.}} \label{tab: 1st table}
\begin{tabular}{c c c c}
\hline
\hline
 & 9 eV & 9.5 eV & 10 eV\\
 \hline
\small{Weighting ratio of different partial waves} & & &\\
\small{$a_0$: $a_1$: $a_2$: $a_3$: }  & \small{1: 0.56: 0.25: 0.45:} & \small{1: 0.42: 0.13: 0.14:} & \small{1: 0.54: 1.03: 0.10:}\\
\small{$b_1$: $b_2$: $b_3$: $b_4$} & \small{0.72: 0.22: 0.56: 0.03}  & \small{0.36: 0.16: 0.22: 0.00} & \small{2.10: 1.15: 0.32: 0.08} \vspace*{.3cm} \\ 
\small{Phase difference ($\Sigma$)} & & &\\
\small{$\delta_{s-p}, \delta_{s-d}, \delta_{s-f}$ (rad)} & \small{3.472, 3.457, 1.486} & \small{3.068, 2.387, 0.310} & \small{3.968, 2.224, 5.036}  \vspace*{.3cm} \\ 
\small{Phase difference ($\Pi$)} & & &\\
\small{$\delta_{p-d}, \delta_{p-f}, \delta_{p-g}$ (rad)} & \small{2.403, 4.956, 4.876} & \small{5.447, 1.48, 0.803} & \small{3.885, 0.0, 5.003}\\ 
\hline
\hline
\end{tabular}
\end{table*}

\section{Results and Discussion}

Figure~\ref{fig:ion_yield} shows the ion yield curve of O$^-$ ions produced from CO due to dissociative electron attachment (DEA) process. The arrows indicate the energies at which the velocity slice images (VSI) are taken. The central sliced images at different electron energies are shown in figure~\ref{fig:3d_pics}. The kinetic energy released (KER) in the process is distributed among the neutral carbon atom and the O$^-$ ion. The kinetic energy distribution of the O$^-$ ions for different incident electron energies are displayed in Figure~\ref{fig:ke_dist}. For 9, 9.5 and 10 eV incident electron energies ions are created with kinetic energy distribution having a single peak near 0 eV. The number of counts gradually decreased to zero near 0.7 eV. But for incident electron energy 10.5 eV onward a second peak appears in the kinetic energy distribution curve. The second peak is located around 0.25 eV for 10.5 eV, around 0.40 eV for 11 eV and around 0.58 eV for 11.5 eV electron energies respectively. All the counts shown in Figure~\ref{fig:ke_dist} are normalized at the zero eV peak. For 9 eV, 9.5 eV and 10 eV incident electron energies the angular distributions of the ions created with kinetic energy between 0 to 0.65 eV are shown on the top of Figure~\ref{fig:ang_dist}. Angular distribution of the ions having kinetic energy in the range between 0 to 0.1 eV, 0 to 0.25 eV and 0 to 0.40 eV for incident electron energies 10.5 eV, 11 eV and 11.5 eV, respectively are shown in the middle of Figure~\ref{fig:ang_dist}. At the bottom of the Figure~\ref{fig:ang_dist} the angular distributions of the ions for 10.5, 11 and 11.5 eV incident electron energies and having kinetic energy between 0.1 to 0.8 eV, 0.25 to 0.65 eV and 0.4 to 1.0 eV respectively are shown. The angular distributions are fitted using different states and four partial waves for each state. According to O'Mallay and Taylor \cite{ang:omalley} the angular distribution of the ions have the general form as  
\begin{equation}
I(k, \theta, \phi) \sim \left|\sum_{L=|\mu|} ^\infty a_{L,|\mu|}(k) Y_{L,\mu} (\theta,\phi)\right|^2 \label{eq:o'mallay}
\end{equation}
due to involvement of each resonant state. We have fitted the angular distribution using equation

\begin{equation}
I(\theta)\sim \sum_{|\mu|}\left|\sum_{j=|\mu|}a_j Y_{j,\mu} e^{i\delta_j}  \right|^2 \label{eq:fit}
\end{equation}

\begin{table*}
\centering
\caption{\small{Fitting parameter for the angular distribution of ions created with lower kinetic energy for incident electron energies 10.5, 11 and 11.5 eV. The angular distributions are fitted for a $\Sigma$ to $\Sigma$ transition only.}}\label{tab:2nd set 1st peak}
\begin{tabular}{c c c c}
\hline
\hline
   & 10.5 eV  & 11 eV  & 11.5 eV\\
   \hline
\small{Weighting ratio of different partial waves} & & &\\
\small{$a_0$: $a_1$: $a_2$: $a_3$} & \small{1: 0.56: 0.17: 0.04} & \small{1: 0.70: 0.26: 0.09} & \small{1: 0.32: 0.15: 0.09} \vspace*{.3cm} \\
\small{Phase difference} & & &\\
\small{$\delta_{s-p}, \delta_{s-d}, \delta_{s-f}$ (rad)} & \small{2.033, 3.231, 4.49} & \small{4.170, 2.922, 1.597} & \small{3.881, 2.105, 0.975}\\
\hline
\hline 
\end{tabular}
\end{table*}

In equation~(\ref{eq:fit}), $\mu=|\Lambda_f - \Lambda_i|$, where $\Lambda_i$ and $\Lambda_f$ are the projection of the electronic axial orbital momentum along the molecular axis for the initial and final molecular states, respectively. The summation over $\mu$ take care of the involvement of the different states in the process. The ground state of neutral CO molecule is $^1\Sigma^+$ ($\Lambda_i=0$). So $\mu$=0, 1, 2 and 3 represents a transition to $\Sigma,\,\Pi,\,\Delta$ and $\Phi$ state respectively, $a_j$'s are the relative weighting factor of the different partial waves, $\delta_j$'s denote the phase differences of the each partial waves with respect to the lowest order partial wave responsible for that particular transition. The potential energy curve calculated by Morgan \emph{et al.} \cite{co:tennyson} showed in Franck-Condon transition region near the resonance energy $\Sigma,\,\Pi,\,\Delta$ and $\Phi$ are present. So the temporary CO$^-$ ion can be formed in any of these states. The angular distribution of the ions for 9 eV, 9.5 eV and 10 eV incident electron energies can be fitted with a single state model for $\Sigma$ to $\Sigma$ transition. Fitting these distribution with a $\Sigma$ to $\Pi$ transition only shows the contribution of $\Pi$ state and the contribution increases with incident electron energies. The best fit is a two state model with a $\Sigma$ to $\Sigma$ and $\Pi$ transition. The angular distributions are fitted with the equation, $|\sum_{j=0} ^3 a_j e^{i\alpha_j} Y_{j,0}|^2 + |\sum_{k=1}^4 b_k e^{i\beta_k} Y_{k,1}|^2$, shown in top of figure~\ref{fig:ang_dist}. Table~\ref{tab: 1st table} shows the parameter used for the best fit to the data. The weighting ratio of the contribution of different partial waves are shown in the first row of the table. The phase difference (in radian) between different partial waves for each states are also shown in the table. Around the 10 eV a b$^3\Sigma^+$ state of CO, as suggested by Sanche and Schulz \cite{co:sanche} might be involved. Comer and Read \cite{co:read} also suggested the presence of this state as a Feshbach resonance. With increasing energy the $^2\Pi$ resonance state near 8 eV shown in figure 2 of \cite{co:tennyson} also gets involved. The angular distribution of the ions for 10.5, 11 and 11.5 eV incident electron energies and having kinetic energy between 0 eV to the first minima value in kinetic energy distribution curve are shown at the middle in Figure~\ref{fig:ang_dist}. These near 0 eV O$^-$ ions are created due to the process II
mentioned in the introduction, having energy threshold of 10.88 eV. Hall \emph{et al.} \cite{co:hall} proposed that the intermediate negative ion state might be a $\Pi$ state. However, our angular distribution data gives the best fit with a $\Sigma$ to $\Sigma$ transition model, using the equation, $|\sum_{j=0} ^3 a_j e^{i\alpha_j} Y_{j,0}|^2$. In Table~\ref{tab:2nd set 1st peak}, the fitted parameters used for the fitting are shown. With increasing energy the contribution from $\Pi$ state increases. The angular distributions show that intensity at 180$^{\circ}$ decreases with increase in incident electron energy. According to Dunn's selection rule \cite{ang:dunn} for heteronuclear diatomic molecule a $\Sigma$ to $\Sigma$ parallel transition has non vanishing probability but $\Sigma$ to $\Pi$ parallel transition has vanishing probability. As the contribution of the $\Pi$ state increases with the increase in incident electron energy the intensity at 180$^{\circ}$ decreases. Individual fitting for $\Pi$ state also shows this increasing contribution. Fitting with individual $\Delta$ and $\Phi$ states shows that they are not contributing in the process. Thus we propose that intermediate state to be mostly $\Sigma$ state with minor contribution from $\Pi$ state. 

The angular distributions of the ions with the higher kinetic energy for 10.5, 11 and 11.5 eV incident electron energies are shown at the bottom of Figure~\ref{fig:ang_dist}. They are attributed due to process I \cite{co:chantry, co:hall} as mentioned in the introduction. The angular distributions had two peaks near 130$^0$ and 230$^{\circ}$, two small lobes around 30$^{\circ}$ and 330$^{\circ}$ and almost no ions in 0$^{\circ}$ but reasonable number of ions along 180$^{\circ}$. The angular distribution has been fitted with four different single state model for a transition to $\Sigma,\,\Pi,\,\Delta$ and $\Phi$ states, and also with multi-state model having different combinations of the states. With a single $\Sigma$ state model the angular distribution gives a good fit with R$^2$ value greater than 0.97. But overestimates the intensity at 180$^{\circ}$ and fails to predict the small lobes around 30$^{\circ}$ and 330$^{\circ}$. A single $\Pi$ state model also depicts the angular distribution reasonably well with R$^2$ value greater than 0.9. The $\Pi$ state model can successfully predicts the two small lobes around 30$^{\circ}$ and 330$^{\circ}$, but gives vanishing intensity at 180$^{\circ}$ as parallel transition from $\Sigma$ to $\Pi$ states is not allowed \cite{ang:dunn}. A $\Sigma$ to $\Delta$ state transition model can also fairly describes the angular distribution but slightly over estimate the intensities around 30$^{\circ}$ and 330$^{\circ}$. This model also gives vanishing intensity at 180$^{\circ}$ as this transition is also forbidden according to Dunn. A $\Sigma \rightarrow \Phi$ state model can also roughly describe the angular distribution, but largely overestimates the the intensities around 30$^{\circ}$ and 330$^{\circ}$ and underestimates at 180$^{\circ}$. Fitting with multi-state models showed that the contribution of $\Phi$ state is vanishingly small and a $\Sigma$ to $\Sigma,\,\Pi$ and $\Delta$ model gives the best fit. A three states model having contribution of four partial waves for each state, of the form $|\sum_{j=0} ^3 a_j e^{i\alpha_j} Y_{j,0}|^2 + |\sum_{k=1}^4 b_k e^{i\beta_k} Y_{k,1}|^2 +|\sum_{m=2}^5 c_m e^{i\gamma_m} Y_{m,2}|^2 $ has been used to fit the angular distribution data. We have not consider any interference between different states as proposed by Tian \emph{et al.} \cite{co:xi}. They have considered the interference effect to minimize the two small forward lobes around 30$^{\circ}$ and 330$^{\circ}$ predicted by the two state model without interference but absent in their experimental data. But surprisingly our experimental result shows the presence of the two small forward lobes. Our model can also predict the forward-backward asymmetry quite well due to the interference between different partial waves involved in the process.
\begin{table*}
\centering
\caption{\small{Fitting parameter for the angular distribution of ions created with higher kinetic energy for incident electron energies of 10.5, 11 and 11.5 eV. The angular distributions are fitted for a $\Sigma$ to $\Sigma$, $\Pi$ and $\Delta$ transition separately.}} \label{tab:2nd_2nd}
\begin{tabular}{c c c c}
\hline
\hline
   & 10.5 eV  & 11 eV  & 11.5 eV\\
\hline
\small{Weighting ratio of different partial waves} & & &\\
\small{$a_0$: $a_1$: $a_2$: $a_3$:} & \small{1: 0.99: 0.82: 1.30:} & \small{1: 0.65: 1.33: 1.68:} & \small{1: 1.19: 0.57: 0.62}\\
\small{$b_1$: $b_2$: $b_3$: $b_4$:} & \small{0.82: 1.67: 1.27: 0.67:} & \small{4.56: 2.91: 1.47: 2.04:} & \small{0.29: 0.73: 0.71: 0.80:}\\
\small{$c_2$: $c_3$: $c_4$: $c_5$} & \small{2.49: 1.16: 0.51: 0.31} & \small{0.34: 2.00: 2.70: 0.30} & \small{3.13: 1.62: 1.21: 0.11} \vspace*{.3cm}\\
\small{Phase difference ($\Sigma$)} & & &\\
\small{$\delta_{s-p}, \delta_{s-d}, \delta_{s-f}$ (rad)} & \small{2.850, 4.002, 0.387} & \small{2.376, 3.457, 0.471} & \small{3.679, 0.225, 1.169} \vspace*{0.3cm}\\
\small{Phase difference ($\Pi$)} & & &\\
\small{$\delta_{p-d}, \delta_{p-f}, \delta_{p-g}$ (rad)} & \small{3.879, 0.866, 2.442} & \small{3.002, 1.779, 4.649} & \small{3.211, 0.361, 2.995} \vspace*{0.3cm} \\
\small{Phase difference ($\Delta$)} & & &\\
\small{$\delta_{d-f}, \delta_{d-g}, \delta_{g-h}$ (rad)} & \small{3.323, 4.261, 4.186} & \small{4.136, 1.207, 1.786} & \small{3.490, 1.389, 1.281}\\
\hline
\hline
\end{tabular}
\end{table*}
In a recent study,  Tian \emph{et al.} \cite{co:xi} reported the angular distributions taken at only two energies, no kinetic energy distributions were reported. The Figure 2 of \cite{co:xi} shows that the central slice images taken at 9.5 and 10 eV gives completely different behavior. We also have observed the similar effect while going from 10 - 10.5 eV. We have studied the ion yield curve (Figure~\ref{fig:ion_yield}) and considered the peak energy to be 9.9 eV \cite{ref:rapp}. We have performed the energy calibration checking before and after taking each set of VSI. Also, above energy different could be due to different energy calibration used in different experiments. Based on measured angular distribution in a limited angular range and considering Dunn's selection rule \cite{ang:dunn}, Hall \emph{et al.} \cite{co:hall} concluded the negative ion resonance  (NIR) state could be a $\Pi$ state. The authors excluded the $\Sigma$ state based on the trend of the experimental finding, i.e. zero counts in $0^{\circ}$ and $180^{\circ}$ directions. However, we are capable to measure the angular distribution over the entire $2\pi$ angle in a very efficient way. We also observed strong forward-backward asymmetry in the angular distribution data, as shown in Figure~\ref{fig:ang_dist}. 

Our data are fitted with the formalism as mentioned above and considering Dunn's selection rule. Here, we conclude the temporary negative ion (TNI) state for the process II to be mainly $\Sigma$ state with little contribution from $\Pi$ state.

\section{Conclusion}
In summary, we have measured the kinetic energy distribution of O$^-$ ions produced from CO due to DEA and the angular distribution of the ions depending on their kinetic energies over the resonance electron energy. Two different dissociation channels are observed with distinct kinetic energies and angular distributions. We do not find any evidence to include the interference effect between different states to describe the angular distribution data. We also conclude that the process I is due to the involvement of $\Sigma$ and $\Pi$ intermediate states, whereas, unlike \cite{co:hall}, we observed main contribution coming from a $\Sigma$ state with minor contribution from $\Pi$ state for the process II.

\begin{small}

\bibliographystyle{unsrt}

\end{small}

\end{document}